\documentstyle[12pt]{article}
\newcommand{\nn}{\nonumber}
\newcommand{\bd}{\begin{document}}
\newcommand{\ed}{\end{document}}
\newcommand{\bc}{\begin{center}}
\newcommand{\ec}{\end{center}}
\newcommand{\be}{\begin{eqnarray}}
\newcommand{\ee}{\end{eqnarray}}
\newcommand{\eqn}{\global\def\theequation}
\newcommand{\sw}{sin^2 \theta_W}
\newcommand{\fbd}{f_B}
\renewcommand{\thefootnote}{\alph{footnote}}
\newcommand{\se}{\section}
\newcommand{\sse}{\subsection}
\newcommand{\bi}{\bibitem}

\def\figcap{\section*{Figure Captions\markboth
     {FIGURECAPTIONS}{FIGURECAPTIONS}}\list
     {Figure \arabic{enumi}:\hfill}{\settowidth\labelwidth{Figure 999:}
     \leftmargin\labelwidth
     \advance\leftmargin\labelsep\usecounter{enumi}}}
\let\endfigcap\endlist \relax
\def\reflist{\section*{References\markboth
     {REFLIST}{REFLIST}}\list
     {[\arabic{enumi}]\hfill}{\settowidth\labelwidth{[999]}
     \leftmargin\labelwidth
     \advance\leftmargin\labelsep\usecounter{enumi}}}
\let\endreflist\endlist \relax

\begin{document}

\tolerance=10000 \baselineskip=7mm 
\begin{titlepage}  

 \vskip 0.5in   
 \null
\begin{center}
 \vspace{.15in}
{\LARGE {\bf  $\tau$ polarization asymmetries in $B^{+}\to
\tau^{+}\nu_{\tau}\gamma$
and $B^{+}_{c}\to \tau^{+}\nu_{\tau}\gamma$
}
}\\
\vspace{1.0cm}
  \par
 \vskip 2.5em
 {\large
  \begin{tabular}[t]{c}
{\bf C.~Q.~Geng, C.~C.~Lih, and T.~H.~Wu}
\\
\\
{\sl Department of Physics, National Tsing Hua University} 
\\  {\sl  $\ $ Hsinchu, Taiwan, Republic of China }\\
   \end{tabular}}
 \par \vskip 5.0em
 {\Large\bf Abstract}
\end{center}

We study the lepton polarization asymmetries in
the radiative leptonic B decays of $B^{+}\to l^{+}\nu_{l}\gamma $ 
and $B^{+}_{c}\to l^{+}\nu_{l}\gamma$.
We concentrate on the transverse component of the $\tau$ lepton asymmetries
in the tau decay modes due to CP violation in theories beyond the standard
model to search T violating effect.

\end{titlepage}

\section{Introduction}

\ \ \

Both CP and T violations have been observed experimentally but so far 
they are only found in the neutral kaon system.
It is known that CP violation implies T violation and vice versa
because of the CPT thoeorem in the local quantum field theory with 
Lorentz invariance and the usual spin statistics,
However, the origin of the CP or T violation is still unclear.
In the standard model, CP violation arises from a unique physical phase 
in the Cabibbo-Kobayashi-Maskawa (CKM) \cite{ckm} quark mixing matrix. To 
ensure the source of CP violation or T violation is this phase indeed, 
we need to consider new processes outside the kaon system. 

In this report, we study the polarization asymmetries of the tau lepton in
$B^{+}\to 
\tau ^{+}\nu _{\tau }\gamma $ and $B^{+}_{c}\to\tau^{+}\nu _{\tau }
\gamma$ decays and focus on the 
transverse parts ($P_T$) of the asymmetries, which are
related to the $T$
odd triple correlation 
\be
P_{T} \propto \vec{s}_{\tau } \cdot (\vec{p}_{\tau } \times 
\vec{p}_{\gamma })\,,
 \ee
where $\vec{s}_{\tau }$ is the tau lepton spin vector and 
$\vec{p}_i\ (i=\tau$ and $\gamma$) 
are the momenta of the $\tau$ and 
photon in the rest 
frame of the $B$-meson.
As in the case of the radiative $K^+_{\mu 2}$ decay \cite{geng1},
the CKM phase does not affect the transverse polarization  in Eq. (1)
and therefore, a non-zero value of $P_T$
could be a signature of physics beyond the standard model.
To illustrate our results,
we will estimate the polarization in some typical non-standard CP violation 
theories, such as the left-right symmetric, three-Higgs doublet, and
supersymmetric models.
The tau transverse polarization for $B^+\to\tau^+\nu_{\tau}\gamma$
could be measured in the two B factories at KEK and SLAC, while that 
for the $B_c^+$ decay could be done at LHC where
about $2\times 10^{8}\ B_c$ mesons are estimated to be produced
\cite{bcpr}.

This report is organized as follows. In section 2, we give a general
analysis for the lepton polarization asymmetries in 
the radiative leptonic 
B decays of interest. We also review the form factors for $B_q\to
\gamma\ (q=u,c)$ transitions calculated directly in the entire physical
range of momentum transfer within the light front framework
\cite{geng2,geng3}. 
In section 3, we concentrate on the transverse part of the tau 
polarization asymmetries in $B_q^+\to \tau^+\nu_{\tau}\gamma$
in various non-standard CP violation models.
Our conclusions are summarized in section~4.

\newpage
\section{Lepton polarization asymmetries in
$B_q^{+}\to l^{+}\nu _l\gamma $}

\ \ \

For a general investigation of lepton polarization asymmetries in 
$B^{+}_{q}\to l^+\nu_{l}\gamma$
including some new CP violating sources, we first carry out the most
general four-fermion interactions given by
\be
{\cal L}=-\frac{G_{F}}{\sqrt{2}}V_{qb}\bar{q}\gamma ^{\alpha }(1-\gamma _{5})b
\bar{\nu }\gamma _{\alpha }(1-\gamma _{5})l +G_{s}\bar{q}b
\overline{\nu }(1+\gamma _{5}) l +G_{P}\bar{q}\gamma_{5}b\bar{
\nu }(1+\gamma_{5}) l  \nn \\ \ +G_{V}\bar{q}\gamma^{\alpha}b\bar{\nu}
\gamma_{\alpha}(1-\gamma_{5}) l+G_{A}\bar{q}\gamma^{\alpha
}\gamma_{5}b\bar{\nu}\gamma_{\alpha}(1-\gamma_{5}) l +H.c.\,,
\label{51}
\ee
where $G_F$ is the Fermi constant, $V_{qb}$ is the CKM mixing element, 
 $q=u$ and $c$, corresponding to $B_u^+\equiv B^+$ and $B^{+}_{c}$
mesons, $l=e,\ \mu$, and $\tau$, and
$G_S$, $G_P$, $G_V$, and $G_A$
denote form factors for the scalar, pseudoscalar, vector, and axial vector
interactions,
arising from new physics,
respectively. 

In the standard model, $G_i\ (i=S,P,V,A)$ vanish and only the first term
in Eq. (\ref{51}) is present.
The decay amplitudes for radiative leptonic B decays of
$B^{+}_{q}\to l^{+}\nu_{l}\gamma$
can be written into two parts
\be
M(B^{+}_{q}\to \l ^{+}\nu _{\l }\gamma )=M_{IB}+M_{SD}\,,
\label{22} 
\ee
where $M_{IB}$ and $M_{SD}$ represent the 
``inner bremsstrahlung'' (IB) and
and ``structure-dependent'' (SD) parts,
and are given by
\be
M_{IB} &=&ie{\frac{G_{F}}{\sqrt{2}}}V_{qb}f_{B_q}m_{l}\epsilon _{\mu
}^{*}B^{\mu }, 
\label{23} 
\\
M_{SD} &=&-i{\frac{G_{F}}{\sqrt{2}}}V_{qb}\epsilon _{\alpha }L_{\beta
}H^{\alpha \beta }\,,
\label{24}
\ee
respectively, with 
\be
B^{\mu } &=&\bar{u}(p_{\nu })(1+\gamma _{5})(\frac{p^{\mu }}{p\cdot q}-%
\frac{2p_{l}^{\mu }+\not{\! }q\gamma ^{\mu }}{2p_{l}\cdot q}%
)v(p_{l},s_{l}), 
\label{25} 
\\
L_{\beta } &=&\bar{u}(p_{\nu })\gamma _{\beta }(1-\gamma _{5})v(p_{l},s_{l}), 
\label{26} 
\\
H^{\alpha \beta } &=&e\frac{F_A^q}{M_{B_q}}(-g^{\alpha \beta }p\cdot
q+p^{\alpha }q^{\beta })+ie\frac{F_V^q}{M_{B_q}}\epsilon ^{\alpha \beta
\mu
\nu }q_{\mu }p_{\nu },
\label{27}
\ee
where $\epsilon_{\mu }$ is the photon polarization
vector, $p$, $p_{l}$, $p_{\nu}$, and $q$
 are the four momenta of $B^{+}_{q}$, $l^{+}$,
$\nu $ and $\gamma$, respectively, $s_{l}$
is the polarization vector of the $l^{+}$, $f_{B_q}$
 is the $B_q$-meson decay constant, and $F^q_{A(V)}$ is the form factor
of the vector (axial-vector) current.
The factors$f_{B_q}$ and $F^q_{V,A}$ are defined as follows:
\be
<0|\bar{q}\gamma ^{\mu }\gamma _{5}b|B_q(p)> &=&-if_{B_q}p^{\mu } \,,
\label{28} 
\\ 
<\gamma (q)|\bar{q}\gamma ^{\mu }\gamma _{5}b|B_q(p+q)>
&=&-e\frac{F_A^q}{M_{B_q}}%
\left[ \epsilon ^{*\mu }(p\cdot q)-(\epsilon ^{*}\cdot p)q^{\mu }\right]
\,,
\label{29}
\\
<\gamma (q)|\bar{q}\gamma ^{\mu }b|B_q(p+q)>
&=&-ie\frac{F_V^q}{M_{B_q}}\epsilon
^{\mu \alpha \beta \gamma }\epsilon _{\alpha }^{*}p_{\beta }q_{\gamma }\,.
\label{210}
\ee
Numerically, $f_{B(B_c)}=0.18\ (0.36)\ GeV$, and
$F^q_{A}$ and
$F^q_{V}$ can be found in the Light-front quark model at the time-like
momentum transfers in which the physical accessible kinematic region is 
$0\leq p^2\leq p^2_{\max }$ with one loop level \cite{geng2,geng3}.

For the IB amplitude, the charged $B_q$-meson  emits leptons 
via the axial-vector
current, and the photon is radiated from the external charged particles
as shown in Figure 1. 
We can calculate the IB part by using the usual rules in QED
 and define it as the ``trivial'' part of the process 
$B^{+}_{q}\to \l ^{+}\nu _{\l }\gamma $. 
We note that the amplitude, $M_{IB}$, is proportional to
the ratio $m_{\ell }/M_{B_q}$ and therefore it is helicity suppressed for
the light charged lepton modes.
The SD amplitude is governed by the vector
and axial vector form factors shown in Figure 2, in which
the photon is emitted from intermediate states.
Gauge invariance leaves only two form factors, $F^q_{V,A}$, 
undetermined in the
SD part.
Since the transverse lepton polarization is deduced by the
interference between IB and SD terms, it is clearly small in the
processes of $B^{+}_{q}\to l
^{+}\nu_{l}\gamma$ with $l=e$ and $\mu$ as the smallness of $M_{IB}$.
We shall focus on the decays of
$B^{+}_{q}\to \tau^+\nu _{\tau}\gamma$ and study the $\tau$
polarization
asymmetries. We remark that the discussions on the decay branching ratios 
of $B^{+}_{q}\to l^{+}\nu _{l}\gamma $ in various models can be found in Refs.
\cite{geng2}-\cite{refbc}.

We now examine the probability of the process $B^{+}_{q}\to \tau
^{+}\nu _{\tau }\gamma $ as a function of the four momenta of the
particles and the polarization vector $s_{\mu }$ of the $\tau $
 lepton. We write the components of $s_{\mu }$ in term of 
$\vec{\eta }$, the unit vector along the $\tau $
 lepton spin in its the rest frame, given by
\be
s_{0}=\frac{\vec{p_{\tau }}\cdot \vec{\eta }}{m_{\tau }%
}\, ,   \vec{s}=\vec{\eta }+\frac{s_{0}}{%
E_{\tau }+m_{\tau }}\vec{p_{\tau }}\,.  
\label{41} 
\ee
In the $B^{+}_{q}$ rest frame, the partial decay rate 
is found to be
\be
d\Gamma =\frac{1}{(2\pi )^{3}}\frac{1}{8M_{B_q}}\mid M\mid ^{2}dE_{\gamma
}dE_{\tau }\,, 
\label{42}
\ee
and
\be
|M|^{2}=A_{0}(x,y)+(A_{L}\vec{e}_{L}+A_{N}\vec{e}_{N}+A_{T}\vec{e}_{T})\cdot 
\vec{\eta}\,,  
\label{43}
\ee
where $\vec{e}_{i}\ (i=L,N,T)$ are the unit vectors along
the longitudinal, normal and transverse components of the tau 
polarization,
defined by
\begin{eqnarray}
\vec{e}_{L} &=&{\frac{\vec{p}_{\tau }}{|\vec{p}_{\tau }|}},  \nonumber \\
\vec{e}_{N} &=&{\frac{\vec{p}_{\tau }\times (\vec{q}\times \vec{p}_{\tau })}{|%
\vec{p}_{\tau }\times (\vec{q}\times \vec{p}_{\tau })|}},  \nonumber \\
\vec{e}_{T} &=&{\frac{\vec{q}\times \vec{p}_{\tau }}{|\vec{q}\times \vec{p}%
_{\tau }|}}\,,  
\label{44}
\end{eqnarray}
respectively. 
The quantities of $A_{0}
$, $A_{L}$, $A_{N}$, $A_{T}$ can be calculated directly and are given by
\begin{eqnarray}
A_{0}(x,y) &=&\frac{1}{2}e^{2}G_{F}^{2}V_{qb}^{2}(1-\lambda )\left\{ 
{\frac{4m_{\tau }^{2}|f_{B_q}|^{2}}{\lambda x^{2}}}\left[
x^{2}+2(1-r_{\tau
})\left( 1-x-{\frac{r_{\tau }}{\lambda }}\right) \right] \right.   \nonumber
\\
&&+M_{B_q}^{4}x^{2}\left[ |F^q_{V}+F^q_{A}|^{2}{\frac{\lambda
^{2}}{1-\lambda }}%
\left( 1-x-{\frac{r_{\tau }}{\lambda }}\right)
+|F^q_{V}-F^q_{A}|^{2}(y-\lambda
)\right]   \nonumber \\
&&-4M_{B_q}m_{\tau }^{2}\left[ Re[f_{B_q}(F^q_{V}+F^q_{A})^{*}]\left(
1-x-{\frac{%
r_{\tau }}{\lambda }}\right) \right.   \nonumber \\
&&-\left. \left. Re[f_{B_q}(F^q_{V}-F^q_{A})^{*}]{\frac{1-y+\lambda
}{\lambda }}%
\right] \right\}\,, 
\label{45}
\end{eqnarray}

\begin{eqnarray}
A_{L}(x,y) &=&-e^{2}G_{F}^{2}V_{qb}^{2}{\frac{(1-\lambda )}{%
2\lambda \sqrt{y^{2}-4r_{\tau }}}}\left\{ {\frac{4m_{\tau
}^{2}|f_{B_q}|^{2}}{%
\lambda x^{2}}}\left[ x(\lambda y-2r_{\tau })(x+y-2\lambda )\right. \right. 
\nonumber \\
&&\left. -(y^{2}-4r_{\tau })(\lambda x+2r_{\tau }-2\lambda )\right]
-M_{B_q}^{4}\lambda x^{2}\left[ |F^q_{V}+F^q_{A}|^{2}{\frac{\lambda
}{1-\lambda }}%
(\lambda y-2r_{\tau })\right.   \nonumber \\
&&\left. \left( 1-x-{\frac{r_{\tau }}{\lambda }}\right)
+|F^q_{V}-F^q_{A}|^{2}\left( y^{2}-\lambda y-2r_{\tau }\right) \right]  
\nonumber \\
&&-4M_{B_q}m_{\tau }^{2}\left[ Re\{f_{B_q}(F^q_{V}+F^q_{A})^{*}\}\lambda
\left( 1-x-%
\frac{r_{\tau }}{\lambda }\right) (2-2x-y)\right.   \nonumber \\
&&\left. \left. +Re\{f_{B_q}(F^q_{V}-F^q_{A})^{*}\}\left( (1-y)(y-\lambda
)+2r_{\tau }-\lambda \right) \right] \right\} \,,   \\
\nn
\\
A_{N}(x,y) &=&e^{2}G_{F}^{2}V_{qb} ^{2}{\frac{(1-\lambda )\sqrt{%
\lambda y-\lambda ^{2}-r_{\tau }}}{M_{B_q}\lambda \sqrt{y^{2}-4r_{\tau
}}}}%
\left\{ {\frac{4m_{\tau }^{3}|f_{B_q}|^{2}}{\lambda x}}(x+y-2\lambda
)\right. 
\nonumber \\
&&-M_{B_q}^{4}m_{\tau }\lambda x^{2}\left[
|F^q_{V}+F^q_{A}|^{2}{\frac{\lambda }{%
1-\lambda }}\left( 1-x-\frac{r_{\tau }}{\lambda }\right)
+|F^q_{V}-F^q_{A}|^{2}\right]   \nonumber \\
&&-2M_{B_q}^{3}m_{\tau }\left[ Re\{f_{B_q}(F^q_{V}+F^q_{A})^{*}\}\left(
{\frac{%
(r_{\tau }-\lambda )(1-x-r_{\tau })}{1-\lambda }}+\lambda x(1-x)\right)
\right.   \nonumber \\
&&\left. \left. -Re\{f_{B_q}(F^q_{V}-F^q_{A})^{*}\}(y-2r_{\tau })\right]
\right\} 
\,,   \\
\nn
\\
A_{T}(x,y) &=&-2e^{2}G_{F}^{2}V_{qb} ^{2}M_{B_q}^{2}m_{\tau }{\frac{%
1-\lambda }{\lambda }}\sqrt{\lambda y-\lambda ^{2}-r_{\tau }}\left\{
Im[f_{B_q}(F^q_{V}+F^q_{A})^{*}]{\frac{\lambda }{1-\lambda }}\right.
\nonumber \\
&&\left. \times \left( 1-x-{\frac{r_{\tau }}{\lambda }}\right)
+Im[f_{B_q}(F^q_{V}-F^q_{A})^{*}]\right\} \,, 
\label{48}
\end{eqnarray}
where $\lambda =(x+y-1-r_{\tau })/x$, $r_{\tau }=m_{\tau
}^{2}/M_{B_q}^{2}$, and $x=2p\cdot q/p^{2}=2E_{\gamma }/M_{B_q}$
 and $y=2p\cdot p_{\tau }/p^{2}=2E_{\tau }/M_{B_q}$ are normalized
energies of the photon and $\tau$, respectively. If we define the
longitudinal, normal and transverse $\tau$ polarization asymmetries by
\be
P_{i}(x,y)={\frac{d\Gamma (\vec{e}_{i})-d\Gamma (-\vec{e}_{i})}{d\Gamma (
\vec{e}_{i})+d\Gamma (-\vec{e}_{i})}}\,,\ (i=L,N,T)\,,
\label{49}
\ee
 we find that
\be
P_{i}(x,y)={\frac{A_{i}(x,y)}{A_{0}(x,y)}}\,,\ (i=L,N,T)\,.
\label{410}
\ee
 From Eqs. (\ref{43}), (\ref{44}) and (\ref{49}), it is easily seen that
the asymmetries of $P_{L}$ and $P_{N}$ are even quantities under
time-reversal transformation, while $P_{T}$ is an odd one. Since
we are interested in T or CP violation, we 
shall give a detail discussion only in the transverse
part of the polarization asymmetries.
Clearly, to have T-odd transverse $\tau$ polarization of $P_{T}$ 
in Eq. (\ref{410}), at least one of the form factors, $f_{B_q}$ and
$F^q_{A,V}$,  has to be complex.
In the standard model, from Eqs. (\ref{48}) and (\ref{410}), 
we see that the
longitudinal and normal $\tau$ polarizations are non-vanishing
but $P_T=0$ since $f_{B_q}$ and $F^q_{V,A}$ are all real
at the tree level. This result is trivial since there is only one diagram 
which contributes to the process of $B^+_q\to \tau^+\nu_{\tau}\gamma$ 
and there is no interference effect or CP violation. At the loop level, 
non-zero value of $P_{T}$ can be induced by electromagnetic final state 
interactions and it is estimated to be $O(10^{-3})$ similar ro the kaon 
case \cite{geng1}. To distinguish the real CP violating effects from the 
final state intereactions at the level of $10^{-3}$, one may measure the 
difference between the tau polarization in $B^+_q\to \tau^+\nu_{\tau}\gamma$
and $B^-_q\to \tau^-\bar{\nu}_{\tau}\gamma$ since the final state 
interactions give the same $P_T$ in both modes whereas CP violation 
yields a different sign for $P_T$.  Beyond the level of $10^{-3}$, an 
observation of $P_T$ in $B^+_q\to \tau^+\nu_{\tau}\gamma$ will indicate T 
violation originated from new physics.

For $B^{+}$ meson, in which $q=u$, using $|V_{ub}|\simeq 3\times 10^{-3}$,
$M_{B}=5.2\ GeV$, $m_{\tau }=1.78\ GeV$, $f_{B}=0.18\ GeV$,
$\tau_{B^+}\simeq 1.62\times 10^{-12}s$ \cite{pdg}, and
$\omega=0.57\ GeV$,
we find that the integrated
branching ratio of $B^+\to \tau^+\nu_{\tau}\gamma$ is given 
by\footnote{Our result here for $B^{+}\to \tau^+\nu_{\tau}\gamma$ agrees 
with those for $B^{+}\to l^+\nu_{l}\gamma\ (l=e,\mu)$ shown in Ref. 
\cite{geng2} after a correction is made. See the note of [19] in Ref. 
\cite{geng3}.}
\be \label{Br}
Br(B^{+}\to \tau^+\nu_{\tau}\gamma) &\simeq & 5.8\times 10^{-7}\,. 
\ee
For the decay of $B^{+}_{c}\to \tau ^{+}\nu _{\tau}\gamma$, 
with
$|V_{cb}|\simeq 4\times 10^{-2}$,
$M_{B_c}=6.4\ GeV$, $f_{B_{c}}=0.36\ GeV$, 
and $\tau_{B^{+}_{c}}\simeq 0.46\times 10^{-12}s$ \cite{bc,cdf}, 
we obtain \cite{geng3}
\be
\label{Brc}
Br(B^{+}_c\to \tau^+\nu_{\tau}\gamma) &\simeq & 1.1\times 10^{-4}\,. 
\ee
 In Figure 3, we show the Dalitz plots of $A_{0}$
for $B_{u,c}^{+}\to\tau^+\nu_{\tau}\gamma$. 
We remark that for the mode $B_q^{+}\to l^{+}\nu _{l}\gamma$ with $l=e$ or
$\mu$,
the transverse lepton polarization asymmetry vanishes in the limit of
$m_l\to 0$ by notifying the lepton mass dependence
in the expression of Eq. (\ref{48}). In such case,
the lepton in the decay is almost $100\%$ longitudinal polarized.
This is why we choose 
$B_q^{+}\to \tau ^{+}\nu _{\tau}\gamma$ to search $P_{T}$
instead of using $B_q^{+}\to \mu^{+}\nu_{\mu }\gamma $.

In the non-standard models, 
$P_T(B_q^{+}\to \tau^+\nu_{\tau}\gamma)$ 
can be non-zero if new physics contains some CP 
violating phases which lead to CP violating physical couplings. These
shall be discussed in the next 
section. However, it is interesting to note that there is no contribution 
to the transverse $\tau$ polarization if the interaction beyond the 
standard model has only left-handed vector current, 
because there is no relative phase between the amplitudes of $M_{IB}$
 and $M_{SD}$.

\section{Transverse $\tau$ Lepton Polarization in Non-standard Theories}

\ \ \ 

It is known that many theories beyond the 
the standard model
provide new CP-violation phases outside the CKM 
mechanism. 
In these theories, sizable $P_{T}$ could be induced.
In this section, we will estimates the values of $P_{T}$ 
in the left-right symmetric,
three-Higgs-doublet, and supersymmetric models, respectively.

  From the interactions in Eq. (\ref{51}), similar to the discussion in 
section 2, we can write the amplitude of the decay 
$B^{+}_{q}\to \tau ^{+}\nu _{\tau}\gamma $ in
terms of the IB and SD contributions. However, 
the factors $f_{B_q}$ and $F^q_{A,V}$ in Eqs. (\ref{23})-(\ref{27}) need
 to be replaced as follows \cite{geng1}:
\be
\nn
f_{B_q} &\to& f_{B_q}(1+\Delta_{P}+\Delta_{A})\,, \\
\nn
F^q_{A} &\to & F^q_{A}(1+\Delta_{A})\,, \\
F^q_{V} &\to&F^q_{V}(1-\Delta_{V})\,,
\label{52}
\ee
with
\be
\Delta_{(P,A,V)}&=&
{\sqrt{2}\over G_F V_{qb}}
\left({G_PM^2_{B_q} \over (m_b+m_q)m_{\tau}},G_A,G_V\right)\,.
\label{eqn:fkp}
\ee
In Eq. (\ref{52}), the factors of $\Delta_{(P,A,V)}$
could be complex numbers due to new physical phases arising from
 $G_{P}, G_{V}, G_{A}$, respectively.
We rewrite $P_{T}(x,y)$ in Eq. (\ref{49}) as
\be
P_{T}(x,y)=P_{T}^{V}(x,y)+P_{T}^{A}(x,y)
\label{54}
\ee
with
\begin{eqnarray}
P_{T}^{V}(x,y) &=&\sigma _{V}(x,y)[Im(\Delta _{A}+\Delta _{V})]\,,  \nonumber
\\
P_{T}^{A}(x,y) &=&[\sigma _{V}(x,y)-\sigma _{A}(x,y)]Im(\Delta _{P})\,,
\label{eqn:pva}
\end{eqnarray}
where
\begin{eqnarray}
& & \sigma _{V}(x,y) =2e^{2}G_{F}^{2}V_{qb} ^{2}M_{B_q}^{2}m_{\tau
}f_{B_q}F^q_{V}{\frac{\sqrt{\lambda y-\lambda ^{2}-r_{\tau }}}{\rho
_{0}(x,y)}}  \nn\\
&& ~~~~~~~~~~ \times \left( {\frac{-1+\lambda }{\lambda }}-
\left( 1-x-{\frac{r_{\tau }}{%
\lambda }}\right) \right) ,  \nonumber \\
& & \sigma _{A}(x,y) =2e^{2}G_{F}^{2}V_{qb} ^{2}M_{B_q}^{2}m_{\tau
}f_{B_q}F^q_{A}{\frac{\sqrt{\lambda y-\lambda ^{2}-r_{\tau }}}{\rho
_{0}(x,y)}}  \nn\\
&& ~~~~~~~~~~ \times \left( {\frac{-1+\lambda }{\lambda }}+
\left( 1-x-{\frac{r_{\tau }}{%
\lambda }}\right) \right) .
\end{eqnarray}

It is clear that, in order to have a non-zero T violating  $P_T$ for
$\tau$, the coupling constants $G_{(P,A,V)}$ have to be present
and furthermore at least one of them is complex to provide the CP
violating phase. In Figures. 4 and 5, we display the
Dalitz plots of $\sigma _{V}(x,y)$ and $\sigma _{V}(x,y)-\sigma
_{A}(x,y)$ in $B^+_{u,c}\to\tau^+\nu_{\tau}\gamma$, respectively. 
  From the figures, we see that they are all
in the order of $10^{-1}$ in most of the allowed parameter space.

\subsection{Left-right symmetric models}

\ \ \
We now study the T violating $\tau $ 
polarization for decay $B_q^{+}\to \tau ^{+}\nu _{\tau }\gamma $ 
in models with 
$SU(2)_{L}\times SU(2)_{R}\times U(1)_{B-L}$
gauge symmetries. In these models, the
Higgs multiplets break the symmetries down to
$U(1)_{em}$ \cite{geng1} and the masses of fermions are generated
from the Yukawa couplings. In addition, the mass eigenstates of gauge
bosons $W_{L,R}$ are related to weak eigenstates, given by 
\be
W_1=\cos\xi \,W_L + \sin\xi \,W_R, \nonumber \\
W_2=-\sin\xi \,W_L + \cos\xi \,W_R,
\ee
where $\xi $ is the left-right mixing angle. Due to the $W_{R}$ gauge
boson, we have the new mixing matrix called the right-handed CKM 
(RCKM) matrix. 
In these models, 
the quark level four-fermion interaction
is given by
\be
{\cal L}_{RL}=-2\sqrt{2}G_{F}(\frac{g_{R}}{g_{L}})K_{qb}^{R^{*}}\xi \bar{b}
\gamma _{\mu }P_{R}q\bar{\nu }\gamma ^{\mu }P_{L}\mu \,,
\label{eq30}
\ee
which contributes $B_q^{+}\to \tau ^{+}\nu _{\tau }\gamma $, 
where $g_{L,R}$ are coupling constants for $SU(2)_{L,R}$ and
$K_{qb}^{R^{*}}$
 is the RCKM matrix element. 
Here, we do not assume parity invariant and any special relation between 
the 
CKM and RCKM matrices. In general, the gauge coupling 
constants $g_{L}$ and $g_{R}$ are unequal to each other
and they are free parameters.
Compare $\cal{L}$ in Eq. (2) with ${\cal L}_{RL}$ in Eq. (\ref{eq30}), 
we find that
\be
G_{A}=G_{V}=-\frac{G_{F}}{\sqrt{2}}(\frac{g_{R}}{g_{L}})K_{qb}^{R^{*}}\xi
\,.
\ee
Because we have the relations between $\Delta _{A,V}$
 and $G_{A,V}$ in Eq. (\ref{eqn:fkp}), we get
\be
\Delta _{A}=\Delta _{V}=-\frac{K_{qb}^{R^{*}}\xi
}{V_{qb}}(\frac{g_{R}}{g_{L}
})\,.
\ee
  From the definition of the $\tau $ transverse
polarization, we obtain
\be
P_{T}=2\sigma _{V}(\frac{g_{R}\xi }{g_{L}})\ Im(K_{qb}^{R^{*}}) .
\ee
In the typical left-right symmetric models
shown in Ref. \cite{lrm1}, one generally has that
\be
\xi\frac{g_{R}}{g_{L}}<4.0\times 10^{-3}\,.
\label{eq34}
\ee
The limit in Eq. (\ref{eq34}) clearly 
leads to a very small $P_T$ even with a large value of 
$Im(K_{qb}^{R^{*}})$.
However, in a class of the specific
models studied in Ref. \cite{lrm2}, it is found
that
\be
\xi {g_R\over g_L} < 3.3 \times 10^{-2}
\label{512}
\ee
for $M_R>549\ GeV$.
With the value in Eq. (\ref{512}) and the assumption of
$2\sigma_{V}Im(K_{qb}^{R^{*}})\sim 1$,
we get that
\be
P_{T}(B_q^+\to\tau^+\nu_{\tau}\gamma)< 3.3\times 10^{-2}\,,
\label{eqn36}
\ee
for both $q=u$ and $c$ modes.
The bound in Eq. (\ref{eqn36}) can be larger if one uses a large value of
$\sigma_{V}$. However, it is hard to be measured in B factories and LHC 
for the $B^+$ and $B_c^+$ decays, respectively.
We remark that similar to the charged kaon case \cite{geng1,gengk1},
the transverse lepton polarizations
for $B_q^+\to (\pi^0,D^0)
l^+\nu_l$ vanish, whereas that for $B_q^+\to D^* l^+\nu_l$ are expected to
be non-zero since both $\gamma$ and $D^*$ are vector particles while
$\pi^0$ and $D^0$ are pseudoscalars.

\subsection{Three-Higgs-Doublet Models}

\ \ \
In the Standard Model (SM), there is no tree level flavor changing neutral
current
 (FCNC) because the fermion masses and Higgs-fermion couplings
can be simultaneously diagonlized. 
In a multi-Higgs doublet model (MHDM), the spontaneous $CP$ 
violation (SCPV) arises from complex vaccum expectation values. 
In the Weinberg three-Higgs-doublet model (THMD) \cite{Weinberg}, 
$CP$ phase occurs in the charged-Higgs-boson mixing and 
it contributes to FCNC at tree level. 
We shall concentrate on this phase and assume that 
the CKM matrix is real in this subsection. 

With the consideration of the natural flavor conservation
(NFC) \cite{Glashow}, the general Yukawa interaction in the models  
is given by
\be
{\cal L}_{Y}=
\bar{Q}_{L_{i}} F^{D}_{ij}\phi_{d} D_{R_{j}}+
\bar{Q}_{L_{i}} F^{U}_{ij}\tilde{\phi} _{u} U_{R_{j}}+
\bar{L}_{L_{i}} F^{E}_{ij}\phi _{e} E_{R_{j}}+h.c.
\ee
with $i,j=1,2,3$,
where $Q_{L_{i}}$, $L_{L_{i}}$ are the left-handed quark and 
lepton doublets, while $U_{R_{j}}$, $D_{R_{j}}$ and 
$E_{R_{j}}$ are right-handed singlets for up, down-type quarks and 
charged leptons, 
$\phi_{k}\ (k=u,d,e)$ are Higgs 
doublets with $\tilde{\phi}_{k}=i\sigma_{2} \phi^{*}_{k}$,
 and  $F_{ij}$ are the coupling constants, respectively.
With fermion mass eigenstates, the Yukawa interaction of physical 
charged scalars is given by 
\be
{\cal L}=(2\sqrt{2}G_{F})^{1/2}\sum_{i=1}^{2}(\alpha_{i}\bar{U}_{L}
KM_{D}D_{R}+\beta_{i}\bar{U}_{R}M_{U}KD_{L}+\gamma_{i}\bar{\nu}_{L}
M_{E}E_{R})H_{i}^{+}+h.c.\,,
\label{eqn:k2}
\ee
where $M_D$, $M_U$, and $M_E$ are the diagonal mass matries of down,
up-type 
quarks and charged leptons, respectively,
$K$ is the mixing matrix for quark sector, $H^+_i\ (i=1,2)$ denote the
two charged physical 
scalars and $\alpha_{i}$, $\beta_{i}$, $\gamma_{i}$ are complex coupling
constants which are related by
\be
\frac{Im\alpha_{1} \beta_{1}^{*}}{Im\alpha_{2} \beta_{2}^{*}}=
\frac{Im\alpha_{1} \gamma_{1}^{*}}{Im\alpha_{2} \gamma_{2}^{*}}=
\frac{Im\beta_{1} \gamma_{1}^{*}}{Im\beta_{2} \gamma_{2}^{*}}=-1\,.
\ee
Comparing Eq. (\ref{eqn:k2}) with Eq. (\ref{51}), we
find that 
the pseudoscalar coupling constant $G_{P}$ has the following form
\be
G_{P}=\sqrt{2}G_{F}V_{qb}m_{\tau}\sum_{i=1}^{2}
\frac{\gamma_{i}}{m_{i}^{2}}(m_{q}\beta_{i}^{*}-m_{b}\alpha_{i}^{*})
\,.
\ee
  From Eq. (\ref{eqn:fkp}), one has that
\be
Im\Delta_{P}=\frac{M_{B_q}^{2}}{m_{b}+m_{q}}\sum_{i=1}^{2}Im
{\frac{\gamma_{i}}{m_{i}^{2}}(m_{q}\beta_{i}^{*}-m_{b}\alpha_{i}^{*})}\,.
\label{eqn41}
\ee
To illustrate the polarization effect, 
we assume that $m_{H_2}\gg m_{H_1}\equiv m_H$.
As shown 
in Refs. \cite{Okada,Grossman}, for $m_{H}<400\ GeV$,
the experimental limit on the inclusive process 
$B\to X\tau\nu_{\tau}$ decay gives the strongest bound
\be
\frac{|Im\alpha_{1}^{*}\gamma_{1}|}{m_{H}^{2}}<0.23\ (GeV)^{-2}\,,
\ee
which leads to
\be
|Im\Delta_{P}|_{u,c} < 5.8 \,,\ 6.7
\label{eqn43}
\ee
where we have ignored the terms relating to $m_q$ in Eq. (\ref{eqn41}).
  From Eq. (\ref{eqn:pva}),
with $|\sigma_V-\sigma_A|\sim 0.1$ and the limit in Eq. (\ref{eqn43}), the
transverse $\tau$ polarization asymmetries of 
$B^{+}_{u,c}\to \tau^{+}\nu _{\tau }\gamma $ 
are found to be
\be
   P_T(B^{+}_{u,c}\to \tau^{+}\nu_{\tau}\gamma)< 0.58\,,\ 0.67\,.
   \ee 
Clearly, a measurement of $P_T$ would also constrain $Im\Delta_{P}$.
We remark that large 
$P_T(B^+\to Ml^+\nu_{l})$ with $M=\pi,D^{(*)}$ for $l=\mu,\tau$ are also
expected \cite{gengt}.

\subsection {Supersymmetric Models without R-parity}
\ \ \ 
It is known that, in general, SUSY theories
would contain couplings with the violation of baryon or/and lepton 
numbers, that could induce the rapid proton decay.
 To avoid such couplings, one usually 
assigns R-parity, defined by $R\equiv (-1)^{3B+L+2S}$ 
to each field, 
where $B(L)$ and $S$ stand for the baryon (lepton) number and 
spin, respectively.
Thus, the R-parity can be used to distinguish the particle (R=+1) from its 
superpartner (R=$-1$).
 In this study, we will discuss the SUSY models 
without R-parity. 
To evade the stringent constraint from proton decay,
we simply require that 
B violating couplings do not coexist
with the L violating ones. 
With the violation of 
the R-parity and the lepton number,
we write the superpotential as
\be
W_{\not\!L}=\frac{1}{2}\lambda_{ijk}L_iL_jE^c_k+\lambda'_{ijk}L_iQ_jD^c_k,
\ee
where the subscripts $i\,,\ j$ and $k$ are the generation indices, $L$ and
$E^c$ 
denote the chiral superfields of lepton doublets and 
singlets, and Q and $D^c$ are the chiral superfields of quark doublets and 
down-type quark singlets, respectively. 
We note that the first two generation indices of
$\lambda_{ijk}$ are antisymmetric, $i.e.$,
$\lambda_{ijk}=-\lambda_{jik}$.   
The corresponding Lagrangian is 
\be
{\cal L}_{\not\! L}&=&\frac{1}{2}\lambda_{ijk}[\bar{\nu}^c_{Li} e_{Lj}
\tilde{e}^*_{Rk}+\bar{e}_{Rk} \nu_{Li} \tilde{e}_{Lj}+
\bar{e}_{Rk}e_{Lj}\tilde{\nu}_{Li}-(i\leftrightarrow j)]
+ \lambda'_{ijk}[\bar{\nu}^c_{Li} d_{Lj}\tilde{d}^*_{Rk} \nonumber \\
&+&\bar{d}_{Rk}\nu_{Li} \tilde{d}_{Rk} 
+\bar{d}_{Rk}d_{Lj}\tilde{\nu}_{Li}- \bar{e}^c_{Ri}u_{Lj}
\tilde{d}^*_{Rk}-\bar{d}_{Rk}e_{Li}\tilde{u}_{Lj}-\bar{d}_{Rk}u_{Lj}
\tilde{e}_{Li}]+h.c..
\label{eqn:lv}
\ee 
 From Eq. ({\ref{eqn:lv}), we find that the four-fermion 
interaction for $b\to q\tau\nu$ with the slepton as the intermediate 
state is given by
\be
{\cal L}_{RV}=-{\lambda^*_{3i3}\lambda'_{ik3}\over M^2_{\tilde{e}_{Li}}}
\bar{b}\,P_L\, q \bar{\nu}\, P_R \, \tau,
\label{eqn:rv}
\ee
where $k=1(2)$ for $q=u(c)$ and $M^2_{\tilde{e}_{Li}}$ is the slepton
mass.

 From the interaction in Eq. (\ref{eqn:rv}), we get
\be
G_P={\lambda^*_{3i3} \lambda'_{ik3} \over 4 M^2_{\tilde{e}_{Li}}}\,,
\ee
which leads to
\be
\Delta_P={\sqrt{2}\over 4G_F V_{qb}}{M^2_{B_q}
\over (m_b+m_q)m_{\tau}}
{\lambda^*_{3i3}\lambda'_{ik3}\over M^2_{\tilde{e}_{Li}}}
\ee
for $i=1,2$.
We therefore obtain the transverse tau polarization of 
$B_q^+\to \tau^+\nu_{\tau}\gamma$ as
\be
 P_T&=&-(\sigma_{V}-\sigma_{A}){\sqrt{2}\over 4G_{F}
V_{qb}}{M_{B_q}^2\over 
(m_b+m_q)m_{\tau}}{ Im(\lambda^*_{3i3} \lambda'_{ik3})\over 
M^2_{\tilde{e}_{Li}}}\,.
\label{eqn:susypt}
 \ee
In order to give the bounds on the R-parity violating couplings, we 
need to examine various processes induced by the FCNC. 
 From the data in Ref. \cite{pdg} on the leptonic $\tau$ decays,
 the bounds on R-parity violating couplings $\lambda$ 
are given by \cite{Barger} 
\be
{|\lambda_{23k}|\over M}&<&4.0\times 10^{-4} {1\over 
GeV}\,.
\label{eqn:b1}
\ee
  From the limit on the electron neutrino mass \cite{pdg}, one has
\cite{nm}
\be
{|\lambda_{133}|\over M}&<&1.0\times 10^{-5}\; {1\over 
GeV}\,.
\label{eqn:b2}
\ee
The bounds on $\lambda'$ can be extracted from the experimental limit on
$K^+\to \pi \nu \bar{\nu}$ as studied in Ref. \cite{Agashe}. One finds
that
\be
{|\lambda'_{ijk}|\over M_{\tilde{d}_{Rk}}}
<1.2\times 10^{-4}\;{1\over GeV}\,,
\label{eqn:b3}
\ee
where $j=1,2$ and
$M_{\tilde{d}_{Rk}}\sim M$=$100$ GeV is the sdown-quark mass. 
 From Eqs. (\ref{eqn:susypt})-(\ref{eqn:b3}), we find
\be
 P_T(B^{+}_{u,c}\to \tau^{+} \nu_{\tau} \gamma)\le 0.16\,,\ 0.02
\ee
where we have used that $|\sigma_V-\sigma_A|\sim 0.1$. 
Similar studies in the various semileptonic B decays have been done 
in Ref. \cite{ng}.

\section{Conclusions}

\ \ \

We have studied the transverse tau polarization asymmetries
in the decays of $B^+_{u,c}\to \tau^+\nu_{\tau}\gamma$ in various CP
violation theories.
We have demonstrated that the asymmetries are zero in the standard model
but they can be large in the non-standard CP violation theories.
Explicitly, we have found that $P_T(B^+_{u(c)}\to \tau^+\nu_{\tau}\gamma)$  
can be large as large as $3.3\times 10^{-2}$ ($3.3\times 10^{-2}$),
$0.58$ ($0.67$), and $0.16$ ($0.02$) in the
left-right symmetric, three-Higgs doublet, and SUSY models.
We note that some of the values could be accessible in the B factories and
LHC. For example, experimentally, to observe the tau transverse
polarization asymmetries with the efficiency of $4\%$ in detecting the 
tau lepton in the three-Higgs doublet models at the
$n\sigma$ level, one needs at least $1.0\times 10^{8}$
($4.0\times 10^5$) $n^2$ $B^+_{u(c)}$ decays. 
Clearly, the measurement of such effect
is a clean signature of CP violation beyond the standard model.

\begin{center}
{\bf ACKNOWLEDGMENTS}
\end{center}

This work was supported in part
by the National Science Council of
Republic of China under contract number NSC-88-2112-M-007-023.

\newpage

\begin{figcap}

\item
Diagrams for ``Inner-bremsstrahlun'' contributions.
\item
Diagrams for ``Structure-dependent'' contributions.
\item
Dalitz plots of $A_{0}(x,y)$ for (a) $B^+\to \tau^+\nu_{\tau}\gamma$
and (b) $B^+_c\to \tau^+\nu_{\tau}\gamma$.
\item
Dalitz plots of $\sigma_{V}(x,y)$ for (a) $B^+\to \tau^+\nu_{\tau}\gamma$
and (b) $B^+_c\to \tau^+\nu_{\tau}\gamma$.
\item
Dalitz plots of $\sigma_{V}(x,y)-\sigma_{A}(x,y)$ for (a) $B^+\to
\tau^+\nu_{\tau}\gamma$ and (b) $B^+_c\to \tau^+\nu_{\tau}\gamma$.

\end{figcap}

\newpage
\begin{figure}[h]
\includegraphics{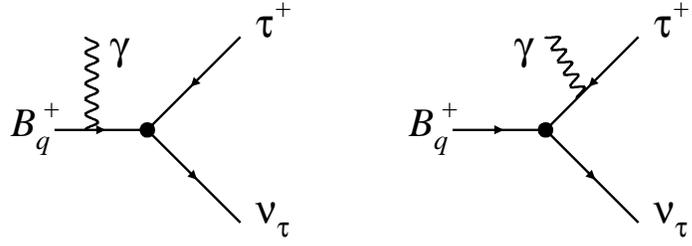}
\vskip 6cm
\caption{Diagrams for ``Inner-bremsstrahlun'' contributions.}
\label{fig:g1}
\end{figure}

\vskip 3cm

\begin{figure}[h]
\includegraphics{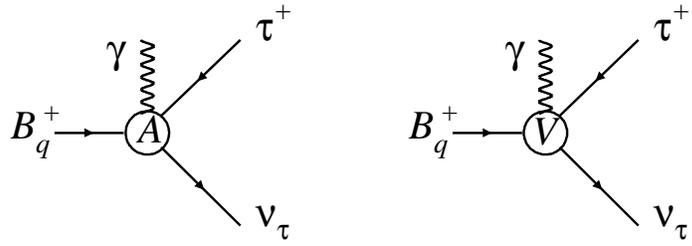}
\vskip 6cm
\caption{
Diagrams for ``Structure-dependent'' contributions.}
\label{fig:g2}
\end{figure}

\newpage
\begin{figure}[h]
\includegraphics{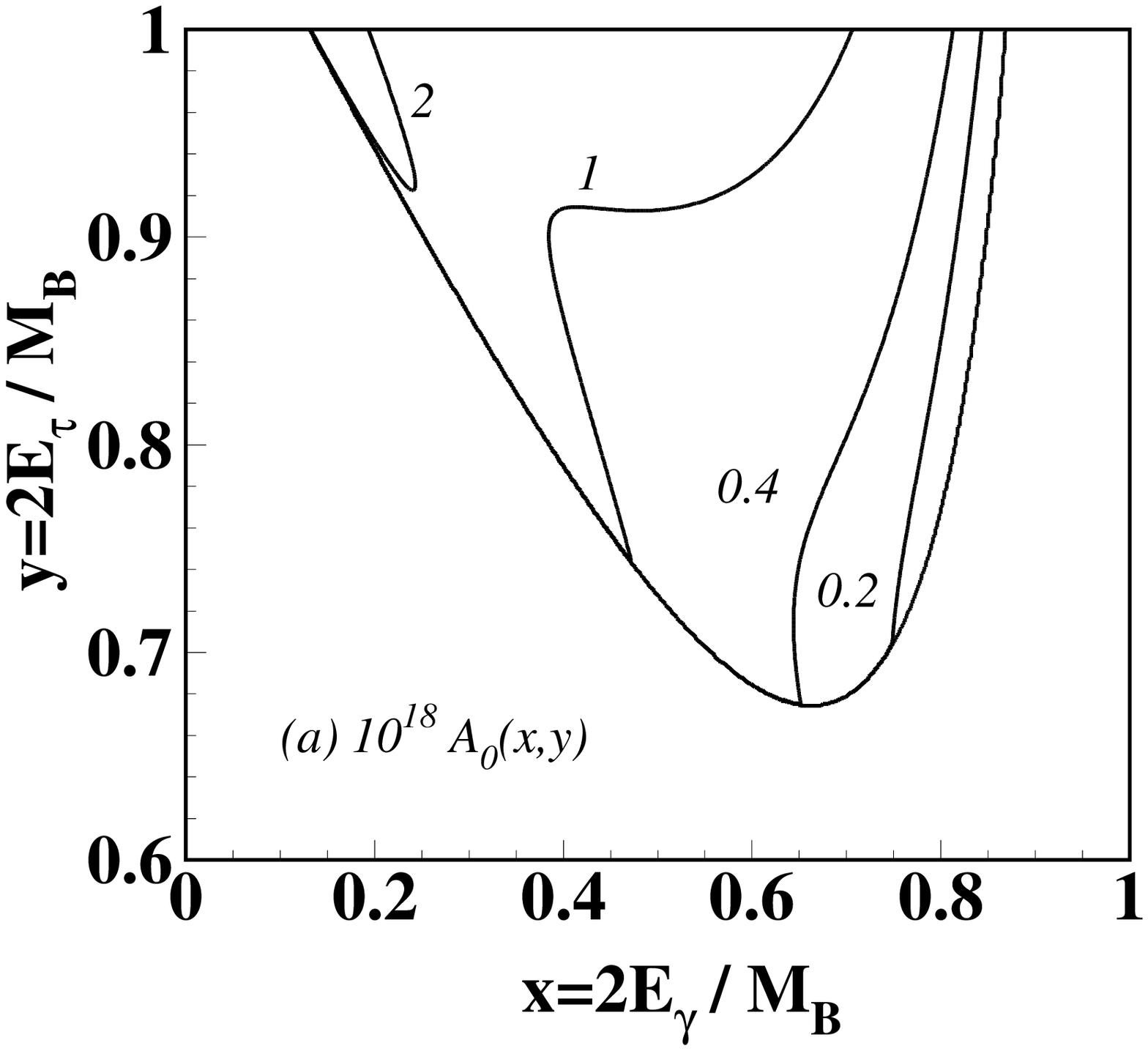}
\vskip 7cm
\end{figure}

\vskip 1.4cm

\begin{figure}[h]
\includegraphics{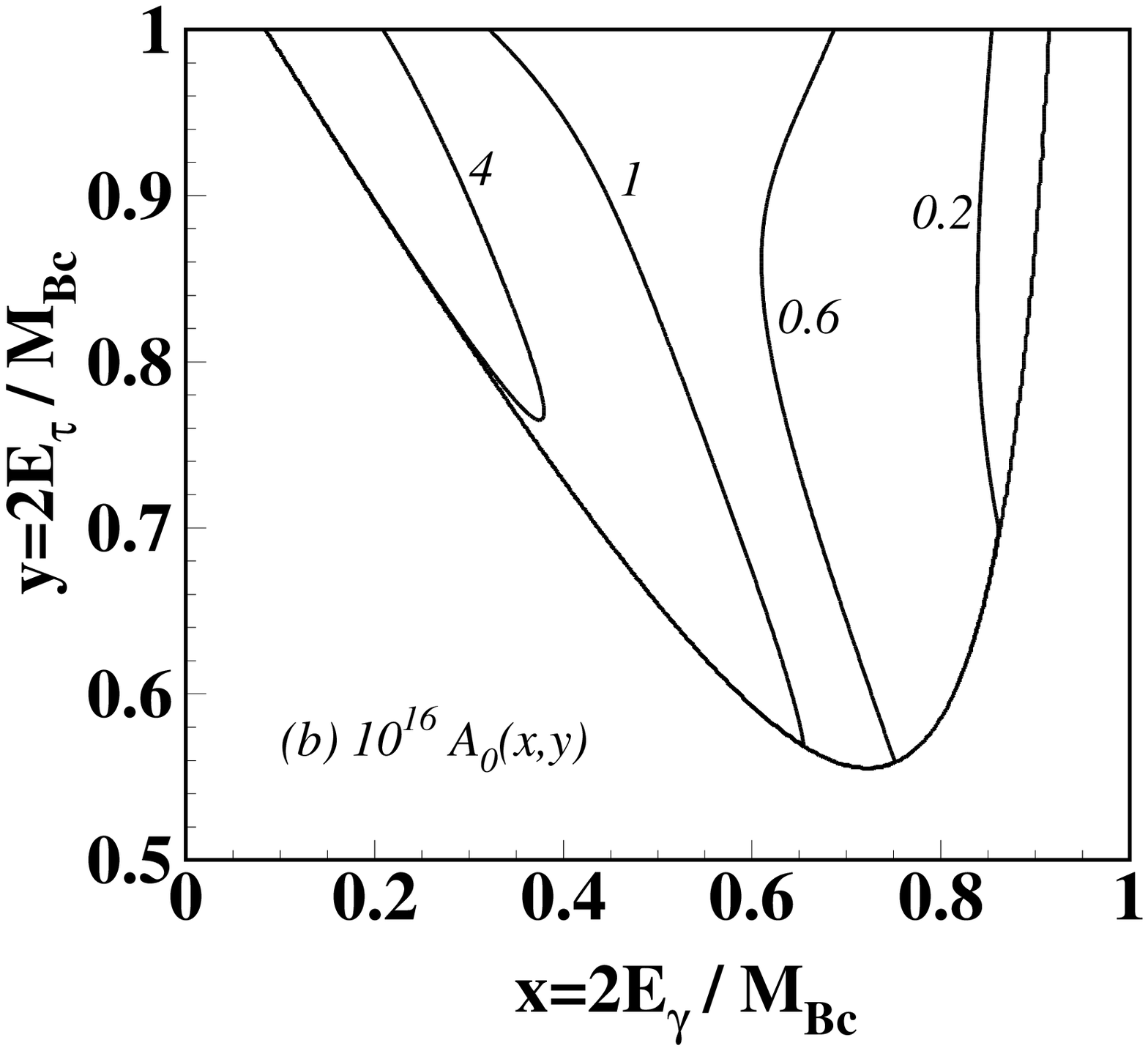}
\vskip 8.8cm
\caption{
Dalitz plots of $A_{0}(x,y)$ for (a) $B^+\to
\tau^+\nu_{\tau}\gamma$ and (b) $B^+_c\to \tau^+\nu_{\tau}\gamma$.}
\label{fig:g3}
\end{figure}

\newpage
\begin{figure}[h]
\includegraphics{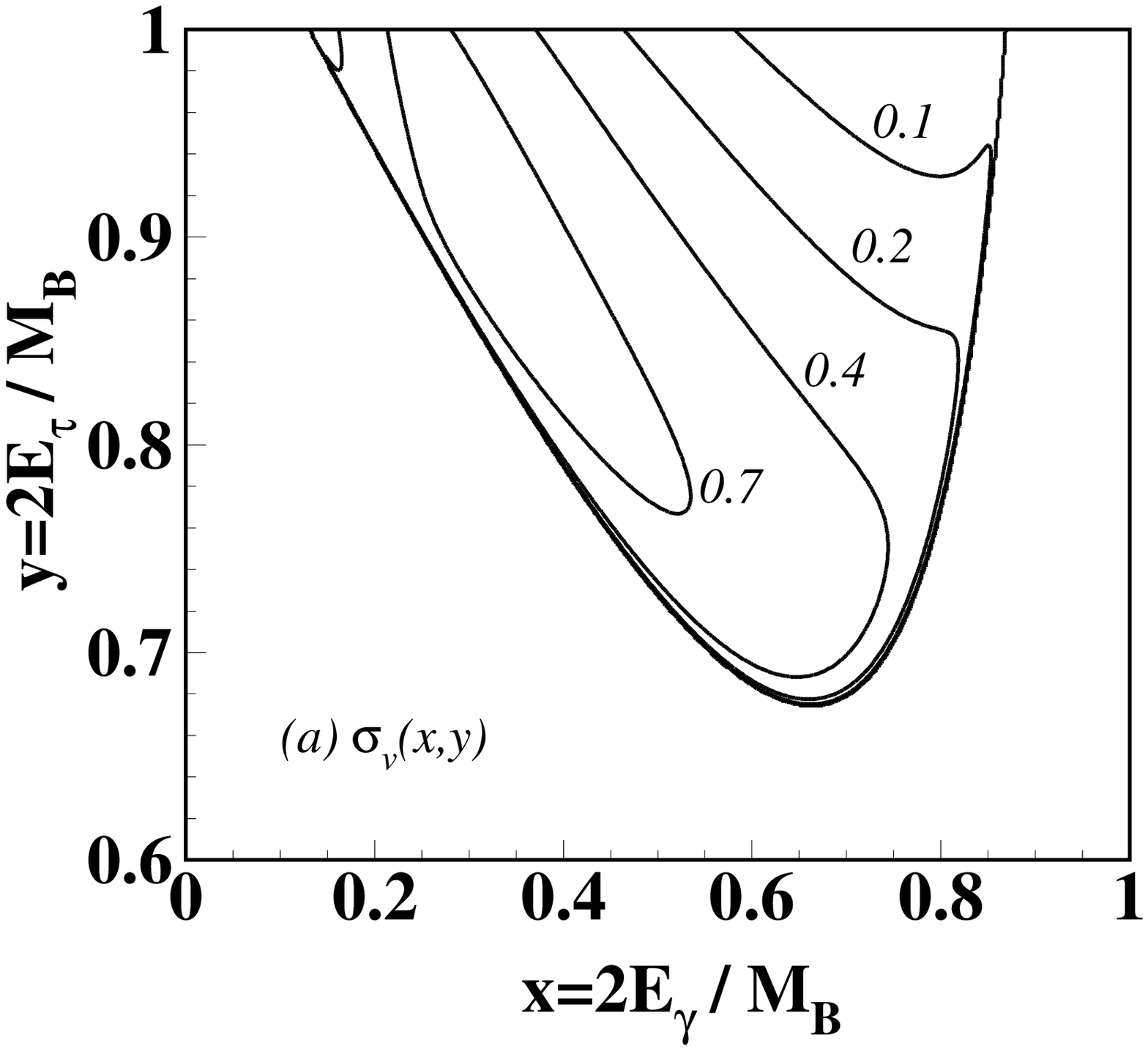}
\vskip 7cm
\end{figure}

\vskip 1.4cm

\begin{figure}[h]
\includegraphics{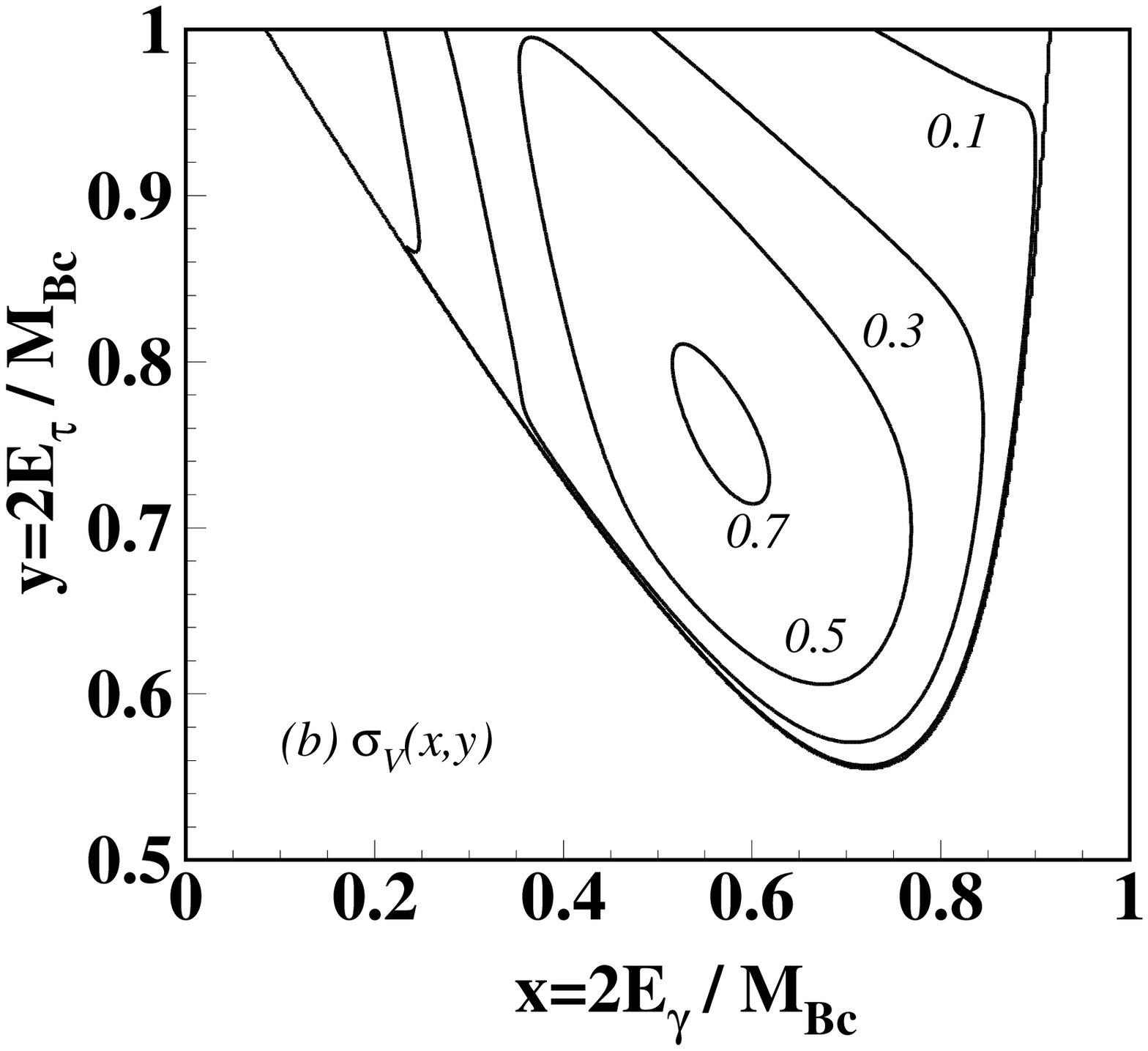}
\vskip 8.8cm
\caption{
Dalitz plots of $\sigma_{V}(x,y)$ for (a) $B^+\to \tau^+\nu_{\tau}\gamma$
and (b) $B^+_c\to \tau^+\nu_{\tau}\gamma$.}
\label{fig:g4}
\end{figure}

\newpage
\begin{figure}[h]
\includegraphics{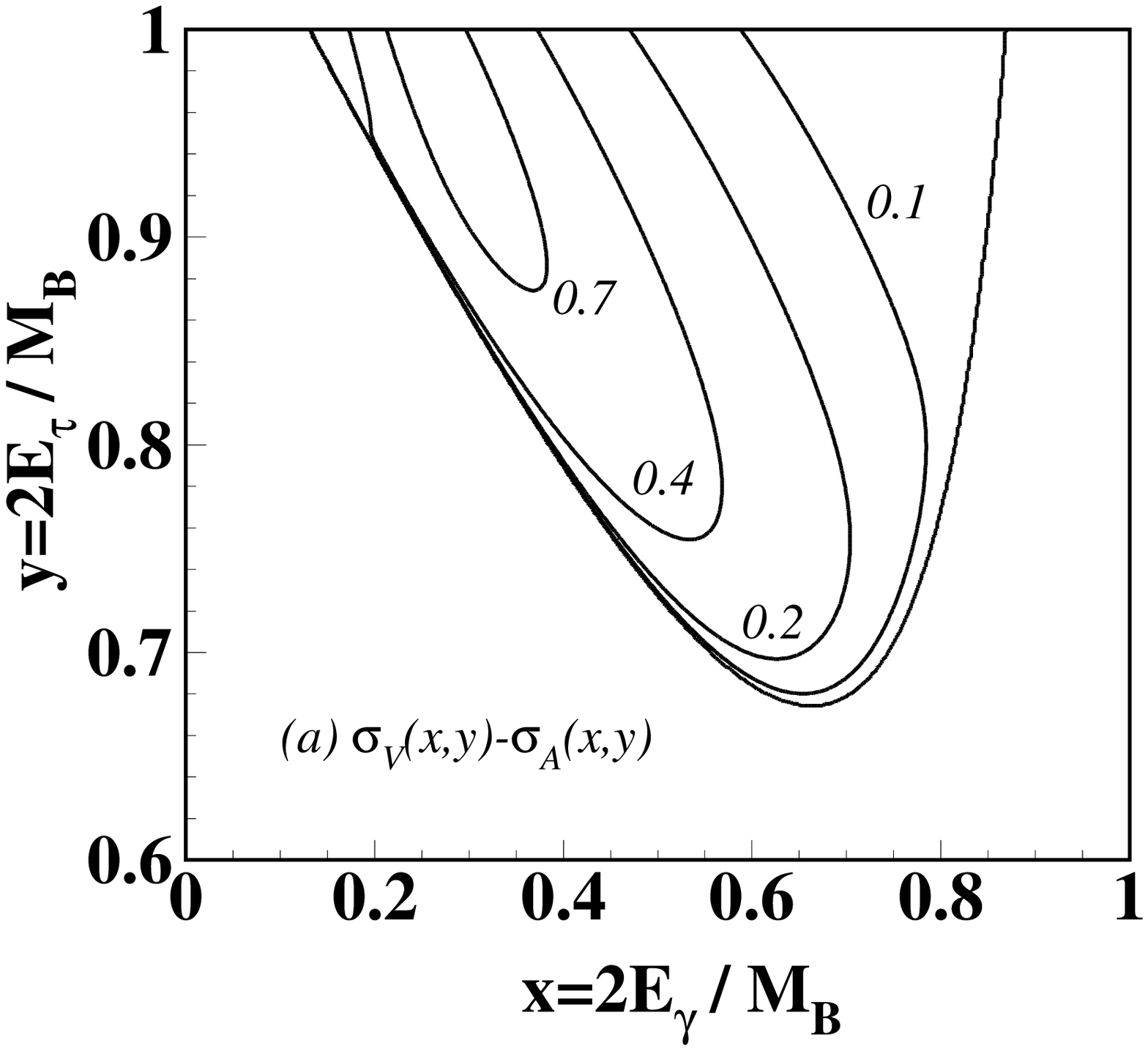}
\vskip 7cm
\end{figure}

\vskip 1.4cm

\begin{figure}[h]
\includegraphics{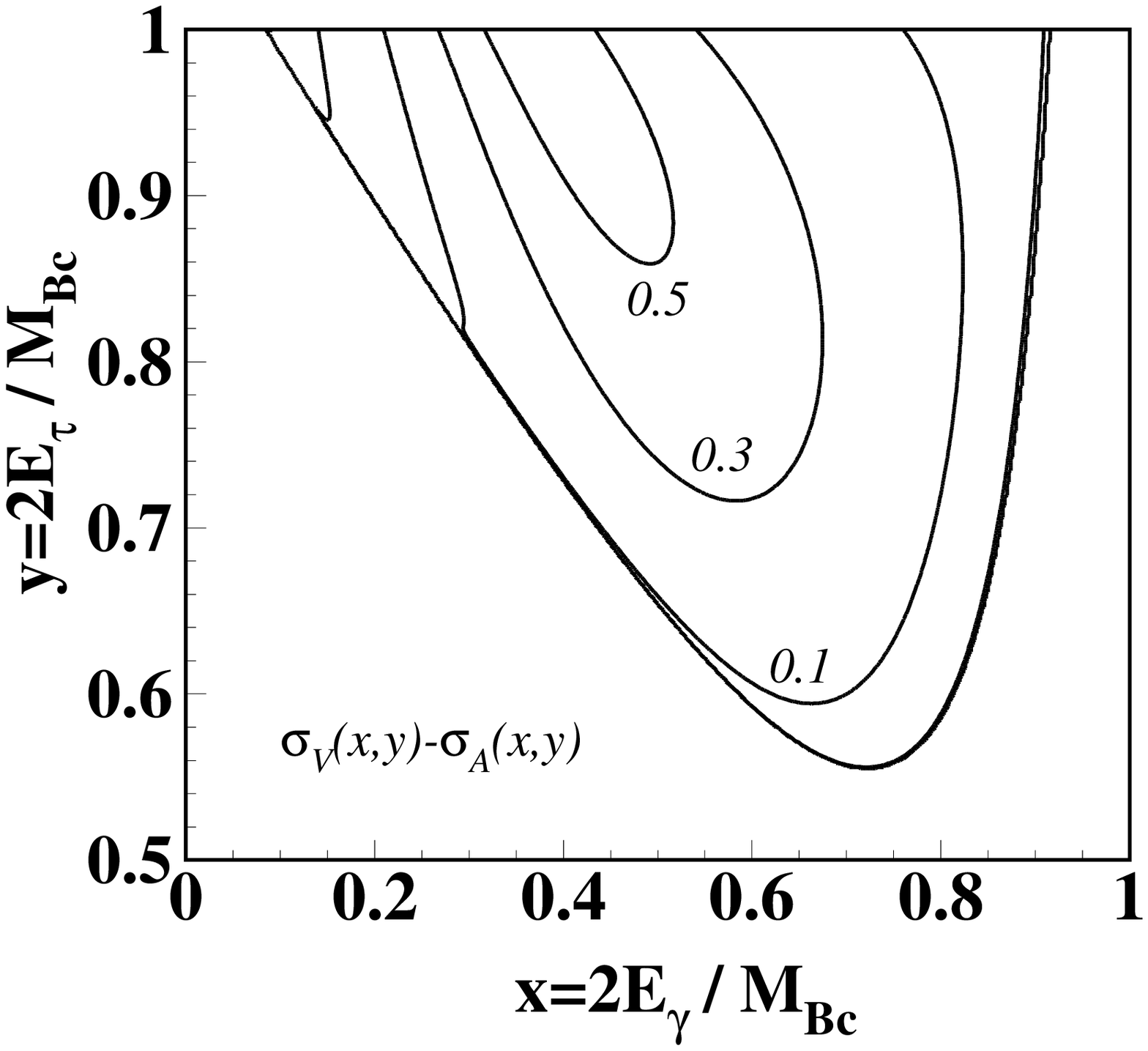}
\vskip 8.8cm
\caption{
Dalitz plots of $\sigma_{V}(x,y)-\sigma_{A}(x,y)$ for (a) $B^+\to
\tau^+\nu_{\tau}\gamma$ and (b) $B^+_c\to \tau^+\nu_{\tau}\gamma$.}
\label{fig:g5}
\end{figure}

\end{document}